# Gold, Oil, and Stocks

Jozef Baruník,[a] Evžen Kočenda,[b] and Lukáš Vácha[c]

Abstract

We employ a wavelet approach and conduct a time-frequency analysis of dynamic correlations between pairs of key traded assets (gold, oil, and stocks) covering the period from 1987 to 2012. The analysis is performed on both intra-day and daily data. We show that heterogeneity in correlations across a number of investment horizons between pairs of assets is a dominant feature during times of economic downturn and financial turbulence for all three pairs of the assets under research. Heterogeneity prevails in correlations between gold and stocks. After the 2008 crisis, correlations among all three assets increase and become homogenous: the timing differs for the three pairs but coincides with the structural breaks that are identified in specific correlation dynamics. A strong implication emerges: during the period under research, and from a different-investment-horizons perspective, all three assets could be used in a well-diversified portfolio only during relatively short periods.

*Keywords*: financial markets, time-frequency dynamics, gold, oil, stocks, high-frequency data, dynamic correlation, financial crisis, wavelets
*JEL Classification*: C01, C13, C58, F37, G11, G15

a Institute of Economic Studies, Charles University, Opletalova 21, 110 00, Prague, Czech Republic and the Institute of Information Theory and Automation, Academy of Sciences of the Czech Republic, Pod Vodarenskou Vezi 4, 182 00, Prague, Czech Republic. E-mail: barunik@utia.cas.cz.

b Corresponding author. CERGE-EI, Charles University and the Czech Academy of Sciences, Politickych veznu 7, 11121 Prague, Czech Republic; CESifo, Munich; IOS Regensburg; The William Davidson Institute at the University of Michigan Business School; CEPR, London; and the Euro Area Business Cycle Network. E-mail: evzen.kocenda@cerge-ei.cz; Phone: (+420) 224005149.

c Institute of Economic Studies, Charles University, Opletalova 21, 110 00, Prague, Czech Republic; Institute of Information Theory and Automation, Academy of Sciences of the Czech Republic, Pod Vodarenskou Vezi 4, 182 00, Prague, Czech Republic. E-mail: vachal@utia.cas.cz.

We benefited from valuable comments we received from Abu Amin, Ladislav Krištoufek, Brian Lucey, Paresh Narayan, Lucjan Orlowski, Perry Sadorsky, Yi-Ming Wei, Yue-Jun Zhang, and participants at several presentations. The support of GAČR grant no. 14-24129S is gratefully acknowledged. The usual disclaimer applies.

# 1 Introduction, Motivation, and Related Literature

Given the extensive spread of the global financial market, co-movements in asset prices receive considerable attention due to their relevance to market integration, portfolio diversification, cross–hedging, and cross-speculation.[1] However, a majority of the empirical analyses that investigate dynamic co-movements employ a time-domain approach that is limited to dynamic links while the frequency analysis of investment horizons is omitted (Ramsey, 2002). Yet, dynamic correlations among assets have been documented to have specific characteristics for particular investment horizons (Conlon et al., 2012), which may be instructive both for policy-makers (financial stability measures) and market participants (predictions of price changes). To better understand the co-movements in asset prices a combined time and frequency analysis is needed. In this paper we take a comprehensive approach to enrich the literature. We perform a time-frequency wavelet analysis of three important assets that have unique economic and financial characteristics: gold, oil, and stocks. For comparison, we also employ standard techniques as well as include a cointegration analysis. By covering a long time span (1987–2012) at both intra-day and daily frequencies and using an array of investment horizons, we deliver a comprehensive study of the dynamic correlations among different classes of major assets.[2] To the best of our knowledge our paper is the first to address the issue of the heterogeneity in correlations among several highly financialized assets over various investment horizons and it brings new insights into their dynamics.

A knowledge of the correlations among assets at different investment horizons is significant for several reasons that are underpinned by the fact that homogenous correlations between assets across investment horizons preclude effective risk diversification in time. First, the importance of various investment horizons for portfolio selection was already recognized by Samuelson (1989). In this respect, Marshall (1994) showed that investors' preference for risk is inversely related to time and different investment horizons have direct implications for portfolio selection.[3] Second, there exist a variety of investors with markedly different investment horizons. Long-term investors base their strategy on fundamental analysis and trade at monthly or yearly horizons. Weekly or daily investors operate on much shorter horizons and base their strategies more on technical analysis. The shortest investment horizons are the domain of speculative traders that operate on an intra-day basis. In such an environment, market activity is necessarily far from homogeneous. Still, the dynamics of the market would be subject to interactions across all trading classes at different investment horizons (Gençay et al., 2010). Third, heterogeneity of market behavior coupled with interactions among assets might result in dynamic correlations among assets that would exhibit less-than-obvious patterns.

---

[1] The literature on co-movements is vast and a full review is beyond the scope of our study. There was a wave of publications on co-movements in top-tier journals in the mid-1990s. More recent contributions include Forbes and Rigobon (2002), Greenwood (2008), Bekaert et al. (2009), Green and Hwang (2009), and Bekaert et al. (2010) with a focus on co-movement factors. Later in this section we introduce studies that are more directly related to our analysis.

[2] The literature on asset co-movements is fragmented in terms of the assets used, time spans, data frequencies, and the techniques employed. Further, correlation analyses usually investigate behavior within a specific class of assets and often disregard the existence of structural shifts.

[3] A parallel can be drawn from the classical term structure theory of interest rates where different maturities are, in a sense, investment horizons as well.

We aim to analyze those patterns simultaneously in the time and frequency domains by employing wavelet analysis. Gençay et al. (2001) and Ramsey (2002) provide ample exposition on the use and versatility of wavelet techniques in economics and finance. During the past decade the methodology gained currency and relevant applications of wavelets include analyses of stocks (Fernandez, 2006, 2008; In and Kim, 2006; Rua and Nunes, 2009), commodities (Vacha and Barunik, 2012; Graham et al., 2013), exchange rates (Nekhili et al., 2002; Karuppiah and Los, 2005; Nikkinen et al., 2011), and other financial and economic variables or their interactions (Kim and In, 2005, 2007; Faÿ et al., 2009; Gallegati et al. 2011; Aguiar-Conraria et al., 2012).

By using wavelets we are able to test the hypothesis on the existence of homogeneity in dynamic correlations across various investment horizons among assets, an issue that so far has been largely overlooked in the literature. In this way we are able to explore the following related questions: To what extent do the assets co-move at different investment horizons? Do correlations among the assets at various investment horizons vary a lot or a little, and are they subject to dramatic changes? Do they share a long-term equilibrium relationship?

For our empirical analysis we chose three assets: gold, oil, and stocks (proxied by the S&P 500). This selection is based on the following reasons: (i) gold and oil represent the most actively traded commodities in the world and the S&P 500 is one of the most actively traded and comprehensive stock indices;[4] (ii) all three assets exhibit marked differences in leverage, which makes them highly interesting from a financial perspective; (iii) there is a motivation for the existence of links among the three assets but empirical evidence is limited to a time-domain approach. Below we review some key facts that further underpin the above reasoning along with some of the literature covering co-movements among the assets under research.

In terms of individual assets, first, gold is traditionally perceived as a store of wealth, especially with respect to periods of political and economic insecurity (Aggarwal and Lucey, 2007). However, gold is a commodity as well as a monetary asset. In this respect Batten et al. (2010) find monetary variables to explain gold volatility. The behavior of gold prices is covered by Lucey et al. (2013). Second, the key importance of oil comes from an industrial perspective and its importance for our society can be documented by the almost 90 million barrels of daily global consumption.[5] As oil is a vital input of production, its price is driven by distinct demand and supply shocks. Lombardi and Robays (2011) find that a short run destabilization in the oil price may be caused by financial investors. However, they argue that while financial activity boosted volatility in the oil market over the recent 2007–2008 crisis, shocks to oil demand and supply remain the main drivers of oil price swings. Over the years, oil also became heavily financialized, as documented in Büyükşahin and Robe (2013). Third, from an economic perspective, stocks reflect the economic and financial development of firms, and market perceptions of a company's standing. They also represent investment opportunities, as well as a link to perceptions of aggregate economic development. Further, stock prices provide helpful information on financial stability and can serve as an indicator of crisis (Gadanecz and Jayaram, 2009). Hence, a wide market index can be used to convey messages on the status and stability of the economy.

---

[4] According to the CME Group Leading Products Resource, S&P 500 futures are traded with the highest average volume among equity indices, gold among metals, and oil among energy commodities (http://www.cmegroup.com/education/featured-reports/cme-group-leading-products.html).
[5] The corresponding consumption figures in millions of barrels daily are 29 for Asia, 18.5 for the U.S.A., and 14.4 for Europe (2012 World Oil Consumption in millions of barrels per day, U.S. Energy Information Administration, assessed on March 1, 2014,
http://www.eia.gov/cfapps/ipdbproject/IEDIndex3.cfm?tid=5&pid=5&aid=2).

From the above account, one may sense that the channels through which the links and co-movements among the assets under research may propagate are not limited only to differences among investors and their investment horizons.[6] For example, the relationship between gold and oil is closely linked to inflation. An inflation channel can well explain the theoretical underpinnings between the two assets: rising oil prices usually influence the aggregate price level (Hunt, 2006) and generate inflationary pressures that prompt hedging against inflation in the form of investments in gold (Narayan et al., 2010). Using a long span of annual data (1960–2005), Baffes (2007) shows that the prices of precious metals, including gold, strongly respond to the price of oil. A similar result is produced by Zhang and Wei (2010), who, based on daily data (2000–2008), find that a rising oil price drives up the price of gold, but they do not find a reverse link.

The above results hint at the potential existence of a long-term equilibrium relationship between gold and oil. Hence, the Efficient Market Hypothesis (EMH) is the basis that motivates an analysis of cointegration among the assets under research. Zhang and Wei (2010) identify a cointegration link between two assets on daily data. Narayan et al. (2010) improve on the cointegration approach and analyze the long-run relationship between gold and oil futures prices over the period 1963–2008 at different levels of maturity in order to gauge differences in hedging behavior. The results indicate that the gold and oil markets are cointegrated, which is presented as evidence of joint market inefficiency. The fact that annual data are employed for the analysis precludes the more detailed and comprehensive results that could be inferred from data of higher frequencies. Studies analyzing the impact of oil prices on stocks across the market show that stock prices rise when the oil price falls and *vice versa* (Huang et al., 1996; Sadorsky, 1999; Faff and Brailsford, 1999). Later studies that analyze the prices of stocks in related (oil, gas) industries show a positive link between those stock prices and the price of oil (Sadorsky, 2001; El-Sharif et al., 2005). Fratzscher et al. (2013) show that oil was not correlated with stocks until 2001, but as oil started to be used as a financial asset, the link between oil and other assets strengthened.

Our key empirical results can be summarized as follows: (i) Correlations between the three assets are low or even negative at the beginning of our sample but following the financial crisis (2007–2009), they dramatically increase. The change in the pattern becomes most pronounced after decisive structural breaks take place (breaks occur during the 2006–2009 period at different dates for specific asset pairs). (ii) Correlations before and after the crisis are homogenous at different investment horizons. (iii) Around the crisis, the heterogeneity in correlations is quite prominent. (iv) Pronounced post-crisis homogeneity in correlations indicates vanishing room for risk diversification based on these assets: until the end of our sample gold, oil, and stocks cannot be used together for risk diversification. (v) In terms of the long-term equilibrium relationship we account for structural breaks and show that the assets under research do not exhibit cointegration.

---

[6] Investment horizons are associated with various types of investors, trading tools, and strategies that correspond to different trading frequencies (Gençay et al., 2010).

The paper is organized as follows. In Section 2, we introduce the theoretical framework of the methodologies we use to perform our analysis. Our large data set is described in detail in Section 3 with a number of relevant commentaries. We bring forth our empirical results in Section 4, in which we present detailed inferences on dynamic correlations as well as long-term relationships among the assets under study. Section 5 concludes.

## 2 Theoretical Framework of the Employed Methodologies

In our study, we employ both standard techniques as well as wavelets. In this section we first provide a brief account and then formalize the techniques. The time domain tools to measure correlations are the nonparametric realized volatility and parametric DCC GARCH methods. While these two approaches are fundamentally different, they both average the relationships over the entire range of possible frequencies and suffer from limited application when dealing with non-stationary time series. When studying market prices with time-series approaches, we have to first-difference the prices to obtain stationary data. This step has economic intuition as first-differenced logarithmic prices are market returns. Still, by this transformation, we lose information about long-term behavior. When analyzing dependencies between asset markets, this may be a crucial loss of information. Wavelets, on the other hand, allow us to analyze time series in the time-frequency domain. Moreover, the wavelet time-frequency domain framework allows for various forms of localization. Thus, when dealing with non-stationary time series, wavelet analysis is better because it is more flexible. Wavelets allow us to simply work with prices and thus study the dynamics of the dependencies not only in time, but also at various investment horizons or frequencies at the same moment. In this way, we may obtain short- as well as long-term dependence structures.

In the following sections we briefly introduce the set of methodologies used for estimating dynamic correlations. We begin with a benchmark parametric DCC GARCH approach, we continue with non-parametric realized measures, and finally introduce an innovative time-frequency approach of wavelet analysis.

### 2.1 Time-varying correlations: The DCC GARCH methodology

Early work on the estimation of time-varying covariances between returns was done by Bollerslev (1990) in his constant correlation model, where the volatilities of each asset were allowed to vary through time but the correlations were time invariant. In a subsequent work, Engle (2002) allowed for dynamics in correlations as well in the now well-established multivariate concept of the Dynamic Conditional Correlation Generalized Autoregressive Conditional Heteroscedasticity (DCC GARCH) model. In this section, we provide a very basic overview of the model.

In Bollerslev's model, correlation matrix $R$ is constant: $H_t = D_t R D_t$, where $D_t = diag\{\sqrt{h_{i,t}}\}$ and $h_{i,t}$ represents the $i$-th univariate (G)ARCH$(p,q)$ process, and $i = 1, ..., n$ where $n$ is a number of assets at time $t = 1, ..., T$. Engle (2002) allowed $R$ to vary in time $t$ thus:

$$H_t = D_t R_t D_t. \qquad (1)$$

The correlation matrix is then given by the transformation

$$R_t = diag(\sqrt{q_{11,t}}, ..., \sqrt{q_{nn,t}}) Q_t diag(\sqrt{q_{11,t}}, ..., \sqrt{q_{nn,t}}), \qquad (2)$$

where $Q_t = (q_{ij,t})$, which is

$$Q_t = (1 - \alpha - \beta)\overline{Q} + \alpha \eta_{t-1} \eta'_{t-1} + \beta Q_{t-1}, \qquad (3)$$

where $\eta_t = \varepsilon_{i,t}/\sqrt{h_{i,t}}$ are the standardized residuals from the (G)ARCH model, $\overline{Q} = T^{-1} \sum \eta_t \eta'_t$ is a $n \times n$ unconditional covariance matrix of $\eta_t$, and $\alpha$ and $\beta$ are non-negative scalars such that $\alpha + \beta < 1$.

To estimate DCC GARCH, we use the standard quasi-maximum likelihood (QML) procedure proposed by Engle (2002) assuming that the innovations are Gaussian. As shown by Engle (2002) and Engle and Sheppard (2001), the DCC model can be estimated consistently by estimating the univariate GARCH models in the first stage and the conditional correlation matrix in the second stage. Parameters are also estimated in stages. This two-step approach avoids the dimensionality problem of most multivariate GARCH models.[7] The above DCC model is parsimonious and ensures that time-varying correlation matrices between stock exchange returns are positive definite.

## 2.2 Time-varying correlations: The Realized Volatility approach

While DCC GARCH estimates the time-varying dynamics of correlations in a parametric way, a simple non-parametric way of estimating the covariance matrix has been developed using high-frequency data. Andersen et al. (2003) and Barndorff-Nielsen and Shephard (2004) suggest estimating the covariance matrix analogously to the realized variation by taking the outer product of the observed high-frequency return over the period. The realized covariance over $[t - h, t]$ for $0 \leq h \leq t \leq T$ is defined as

$$\widehat{RC}_{t,h} = \sum_{i=1}^{M} \mathbf{r}_{t-h+\left(\frac{i}{M}\right)h} \mathbf{r}'_{t-h+\left(\frac{i}{M}\right)h}, \qquad (4)$$

where $M$ is the number of observations in $[t - h, t]$. Details can be found in Andersen et al. (2003) and Barndorff-Nielsen and Shephard (2004), who show that the *ex-post* realized covariance $\widehat{RC}_{t,h}$ is an unbiased estimator of the *ex-ante* expected covariation. Moreover, with increasing sampling frequency the realized covariance is a consistent estimator of the covariation over any fixed time interval $h > 0$ as $M \to \infty$. In practice, we observe only discrete prices, thus bias from discretization is unavoidable. Much more damage is caused by market microstructure effects such as price discreteness, bid-ask spread, and bid-ask bounce. Hence, when using this estimator in practice, one is left with the advice not to sample too often. While the optimal sampling frequency resulting from the vast research on the noise-to-signal ratio[8] can be used, this approach still causes a large amount of available data to be discarded. The best trade-off between reducing bias and losing information is to use the standard 5-minute sampling frequency as suggested by Andersen and Benzoni (2007). We follow their approach and use 5-minute data for the calculation of realized covariances. One last important assumption about the process is that the data are assumed to be synchronized: this means that the prices of the assets have to be collected at the same time. This is not an issue here as in our analysis all three assets under research are paired under equal-stamps matching.

---

[7] In this respect, Bauwens and Laurent (2005) show that both the one-step and two-step methods provide very similar estimates.

*2.3 Time-frequency dynamics in correlations: The Wavelet approach*

While both DCC GARCH and realized volatility approaches allow us to study the covariance matrix solely in the time domain, we are interested in studying its time-frequency dynamics. In other words, we are interested to see how the correlations vary over time and over different investment horizons.

Wavelets offer a decomposition of the economic relationship into time-frequency components. In wavelet analysis, scale is often used instead of frequency, because scale usually represents broader frequency bands. The set of scales represents investment horizons (resolution levels) at which we can study the relationships separately, i.e., on a scale-by-scale basis (Gallegati et al., 2011). Each scale therefore describes the time development of the economic relationship at a particular scale but also dynamically in time. The wavelet decomposition usually provides a broader picture when compared to the time domain approach that in fact aggregates all investment horizons together. Hence, if we expect that economic relationships follow different patterns at various investment horizons, then the wavelet decomposition can reveal interesting characteristics of the data that would otherwise remain hidden.

In our analysis we use a discrete version of wavelet transform called maximal overlap discrete wavelet transformation (MODWT). This transform is a translation-invariant type of discrete wavelet transformation, i.e., it is not sensitive to the choice of the starting point of the examined process. Furthermore, the MODWT does not use a downsampling procedure, therefore the wavelet and scaling coefficient vectors, which are the outcomes of the decomposition, have equal length at all scales similar to the number of observations of the decomposed time series. As a consequence, the MODWT is not restricted to sample sizes that are powers of two.

To obtain the MODWT wavelet and scaling coefficients from time series $x(t)$, we use the pyramid algorithm (Percival and Walden, 2000). A vector of wavelet coefficients is denoted as $W_x(j,s)$, where $j$ denotes the scale and parameter $s$ represents the position that corresponds to the time position of the decomposed time series $x(t)$. For time series $x(t)$, $t = 1,2,...,N$, we obtain $j = 1,...,J$ vectors of wavelet coefficients, where $J \leq log_2(N)$ represents the maximum level of the wavelet decomposition. Generally, the $j$-th level wavelet coefficients in vector $W_x(j,s)$ represent a frequency band $f \in [1/2^{j+1}, 1/2^j]$, whereas the $j$-th level scaling coefficients in vector $V_x(J,s)$ represents frequency band $f \in [0, 1/2^{j+1}]$. Hence, as we increase the number of scales only, the scaling coefficients vector represents a small portion of the spectra. For example, at the first scale $j = 1$, representing the lowest scale (highest frequency), we obtain vector $W_x(1, s)$; in case we have 5-minute data, the first scale represents activity at investment horizons of 10-20 minutes. The second scale coefficients, $W_x(2, s)$, represent investment horizons of 20–40 minutes. The full wavelet decomposition of time series $x(t)$ results in a set of vectors of wavelet and scaling coefficients: $W_x(1, s), W_x(2, s), \ldots, W_x(J, s), V_x(J, s)$. Since the MODWT is an energy preserving transform, the variance of the decomposed time series $x(t)$, $t = 1, 2, \ldots, N$ is completely encompassed in the wavelet scaling coefficients:

$$\|x\|^2 = \sum_{j=1}^{J} \sum_{s=1}^{N} \|W_x(j, s)\|^2 + \sum_{s=1}^{N} \|V_x(J, s)\|^2. \tag{5}$$

---

[8] The literature is well surveyed by Hansen and Lunde (2006), Bandi and Russell (2006), McAleer (2008), and Andersen and Benzoni (2007).

For a more detailed introduction to wavelets, see Daubechies (1988), Percival (1995), and Percival and Walden (2000).

Finally, wavelet correlation allows for an alternative study of dependence between two time series. As wavelets decompose the time series on a scale-by-scale basis, we can estimate correlations for various time horizons represented by scales.

As a first step, a wavelet transform is performed on the examined time series, i.e., the prices of gold, oil, and stocks. As an output from the wavelet transform we obtain vectors of wavelet and scaling coefficients. Since we use the maximal overlap discrete wavelet transform (MODWT), all vectors of wavelet coefficients have the same length.

Following Whitcher et al. (2000), we define wavelet correlation $\rho_{xy}(j)$ between time series $x$ and $y$ at scale $j$ as:

$$\rho_{xy}(j) = \frac{cov(W_x(j,s), W_y(j,s))}{[var(W_x(j,s)) var(W_y(j,s))]^{\frac{1}{2}}} \equiv \frac{\gamma_{xy}(j)}{v_x(j) v_y(j)}, \tag{6}$$

where $W_x(j, s)$ and $W_y(j, s)$ are vectors of the MODWT wavelet coefficients for time series $x(t)$ and $y(t)$ at scale $j$. For instance, using 5-minute data, the wavelet correlation at scale $j$ describes the correlation at an investment horizon of 10–20 minutes. We provide the details about the wavelet variance $v_x^2(j)$ and wavelet covariance $\gamma_{xy}(j)$ in Appendix A and Appendix B. Using the definition of the wavelet correlation (6) we can write an estimator of the wavelet correlation in the form:

$$\hat{\rho}_{xy}(j) \equiv \frac{\hat{\gamma}_{xy}(j)}{\hat{v}_x(j) \hat{v}_y(j)}, \tag{7}$$

where $\hat{\gamma}_{xy}(j)$ is the estimator of wavelet covariance at scale $j$ and $\hat{v}_x(j)^2$ and $\hat{v}_y(j)^2$ are estimators of wavelet variance and covariance, respectively. The central limit theorem for estimator (7) was established by Whitcher et al. (1999). Approximate confidence intervals for the MODWT wavelet correlations are constructed based on Whitcher et al. (1999); empirical values are reported in Section 4.1.

## 3  Data

For our analysis, we use the prices of gold, oil, and the broad U.S. stock market index S&P 500. The data set consists of the tick prices of gold, oil, and S&P 500 futures traded on the platforms of the Chicago Mercantile Exchange (CME) and obtained from Tick Data, Inc.

More specifically, oil (Light Crude) is traded on the New York Mercantile Exchange (NYMEX) platform, gold is traded on the Commodity Exchange, Inc. (COMEX), a division of NYMEX, and finally S&P 500 is traded at the CME in Chicago. We use the most active rolling contracts from the pit (floor traded) session. The prices of all futures are expressed in U.S. dollars.

The sample period spans from January 2, 1987 until December 31, 2012. We acknowledge the fact that the CME introduced the Globex(R) electronic trading platform on Monday, December 18, 2006, and began to offer nearly continuous trading. However, we restrict the analysis of intraday 5-minute returns within the business hours of the New York Stock Exchange (NYSE) as most of the liquidity of S&P 500 futures comes from the period when U.S. markets are open. The time synchronization of our data is achieved in such a way that gold and oil prices are paired with S&P 500 by the same Greenwich Mean Time (GMT) stamp. We eliminate transactions executed on Saturdays and Sundays, U.S. federal holidays, December 24 to 26, and December 31 to January 2, because of the low activity on these days, which could lead to estimation bias. Hence, in our analysis we work with data from 6472 trading days. In Table 1 we present the descriptive statistics of the returns of the data that constitute our sample. The distributions of intra-day as well as daily returns are standard, although a very high excess kurtosis of 104.561 for oil can be noted. This is caused by a single maximum value of a 16.3% return on January 19, 1991, the day when the worst intentional environmental damage ever was caused by the late Iraqi leader Saddam Hussein ordering a large amount of oil to be spilled into the Persian Gulf (for an assessment see Khordagui and Al-Ajmi, 1993). We show the development of the prices of the three assets in Figure 1.

## 4 Empirical Analysis of Gold, Oil, and Stocks relationships

### 4.1 Time-varying correlations

We present dynamic correlations for each pair of assets graphically in Figures 2–4. Each figure contains two panels that plot correlations obtained by the three methods described in Section 2. In the upper panel, the figures report realized volatility-based correlations computed on 5-minute returns for each day and daily correlations from the DCC GARCH(1,1) estimates. The lower panels contain time-frequency correlations computed using wavelet decomposition of 5-minute data.[9] We depict only four investment horizons as examples: 10 minutes, 40 minutes, 160 minutes and 1.6 days.

There are several key features we can infer from the plots. Correlations for asset pairs exhibit stable and similar patterns—correlations are low or even negative—until 2005 between gold and oil, until 2001 between gold and stocks, and until 2004 between oil and stocks. After these years the pattern of correlations between assets radically changes. Dynamic correlations between pairs of variables exhibit the same patterns but their plots differ in the extent of detail depending on what method is used. Correlations based on realized volatility provide very rough evidence. More contoured correlation patterns are inferred from the DCC-GARCH method. The wavelet approach shows its advantage over the two other methods as it offers individual correlation patterns for a number of investment horizons. Thus, it provides true time-frequency research output.[10]

---

[9] For improved clarity of the time-frequency plot, we report aggregated monthly correlations.

We now turn to more detailed inferences. Intraday correlations for the gold-oil pair are remarkably low during the period 1992–2005 at short (10-minute) as well as longer horizons (this result is more accessible from Table 2, presented in the next section). Higher correlations during the period 1990–1991 correspond to the spike visible in graphical form (Figure 2) that is associated with the economic recession in the U.S. (July 1990 to March 1991). A significant increase in correlation begins in 2006. Contrary to only a temporary increase in correlation during 1990–1991, the correlation structure between gold and oil has changed from the recent financial crisis until the end of our sample. This is an interesting result that indicates the existence of a decisive structural break in the correlation structure between the two assets. The result is actually corroborated by the evidence of the structural break on September 8, 2006 reported in Section 4.3.

The presence of negative correlations is a feature that is common to the gold-stocks and oil-stocks pairs (Figures 3 and 4). Negative correlations are actually frequent for the two pairs but they occur more often for the gold-stocks pair (Figure 3). The period 1991–1992 is characterized by negative correlations, especially at longer horizons, and specifically the gold-stocks pair has very rich correlation dynamics since 2001: the correlations reached their minima during 2002–2003, then a steady increase followed. From 2006 this pair has significantly high correlation, except for two short periods in 2008 and 2009. Contrary to the other two pairs, in 2012 there was a very significant increase in the correlation between gold and stocks at all scales. We can observe an increase in the magnitude that is three times larger relative to the previous year. Future research will show whether 2012 was only an anomalous year or whether we are witnessing a more profound change in the correlation structure.

The pair oil-stocks also exhibits increased correlation after the onset of the financial crisis (Figure 4). However, unlike the other two pairs, the correlation between oil and stocks before the crisis was significantly lower than after the crisis. This indicates that the correlation structure of this pair was strongly influenced by developments in 2008. From 2009 on, the oil-stocks pair has the highest correlation of the three examined pairs. Moreover, often we observe highly similar correlation on all scales.

*4.2 Homogenous vs. heterogeneous correlations*

Figures 2–4 provide important insights into the correlation dynamics but cannot offer more precise inferences. Therefore, we proceed with a detailed test of our hypothesis of homogeneous correlations across investment horizons (within each year). Formally we can write the hypothesis as $H_0$: $\hat{\rho}_{xy}(j) = \hat{\rho}_{xy}(i)$ for $i,j \in \{1,2,3,4\}$, where $j$ and $i$ are wavelet scales representing investment horizons, and $i \neq j$. We test this hypothesis on high-frequency data for each year. The hypothesis allows testing whether there is heterogeneity in correlations that is theoretically underpinned by differences in investors, their beliefs, investment strategies, and market links among assets.

---

[10] Despite the fact that the wavelet approach is superior to the other two methods in terms of dynamic correlation analysis, we employ the other methods for the purpose of comparison and completeness of analysis as these are benchmark methods.

We summarize our results in Tables 2–4. First, for each pair of variables, we present an individual table containing the summarized correlations over the period of one year. Each table contains two sets of results where correlations are computed on various investment horizons obtained by the wavelet decomposition for each year. The first set of correlations is based on high-frequency intraday data for different investment horizons ranging from 10 minutes ($j = 1$) to 80 minutes ($j = 4$). The second set contains daily correlations with investment horizons ranging from 2 days ($j = 1$) to 32 days ($j = 5$). For both data resolutions, we also offer a low-frequency component (about one year; columns labelled 160 min-year and 64 day-year). In order to support the interpretation of our results we compute confidence bands around the reported point estimates. The 95% confidence intervals of the correlation estimates are nearly symmetric with maximum values ranging from ± 0.014 for the first scale to ± 0.04 for the last scale.[11] The years in which the hypothesis of homogeneous correlations across scales (investment horizons) is rejected within the 95% confidence interval are highlighted.

Our key finding on the intraday frequency is that in a uniform way we are able to reject the hypothesis of homogeneity in correlations for all three pairs during specific periods that are common to all three pairs. With some minor gaps in years we reject the hypothesis during the early years of our sample (1987–1991), and then immediately prior to and during the financial crisis (2006–2009). Hence, periods of heterogeneity in correlations for all three pairs are present around prominently troubled times: first, the 1987 market crash slowly turning into the early 1990s recession, and second, the financial crisis in 2007–2008. During the rest of our data span the correlations are quite homogenous for the gold-oil and oil-stocks pairs, while the gold-stocks pair exhibits substantial heterogeneity in correlations over time.

Daily correlations offer a more complex pattern. The correlations change their values at various investment horizons more often than the intraday-based correlations. Further, both periods of increased correlations, in the early 1990s and around the financial crisis, are present but they do not stand out so sharply because during other years correlations at specific investment horizons are also quite high. Finally, daily correlations are homogenous across the chosen horizons since the hypothesis could not be rejected. Correlations seem to be higher at longer investment horizons of about one month, while they are very low at short investment horizons measured in days. From 2006 the pattern changes, though. First, correlations alter their magnitudes quite often and these changes are unparalleled in the past. Second, markets become quite homogeneous in the perception of time. Correlations at shorter and longer investment horizons become less diversified. This indicates that differences in short and long investment horizons diminish. A greater homogeneity in the perception of these differences among investors may reflect a heightened uncertainty on markets as well as questionable economic performance in many developed countries following the crisis.

Intraday-based correlations provide further insights (Tables 3–4). First, the results show an overwhelming evidence of heterogeneity in correlations between gold and stocks (Table 3). With some gaps for individual years, we are able to reject the hypothesis of homogeneous correlations for most of the sample from 1987 to 2007. Second, in terms of heterogeneity, the results of the oil-stocks pair (Table 4) resemble the outcome for the gold-oil pair (Table 2). In the early years of our sample we also witness a prevailing heterogeneity in the correlations across horizons during troubled times (1988, 1990–1991). Another period of heterogeneity is visible during the crisis period (2008–2009); in this sense we find heterogeneity in correlations later than in the case of the gold-oil pair.

---

[11] To conserve space, we do not report confidence intervals for all estimates. The results are available upon request from the authors.

Further, an interesting pattern emerges in the heterogeneity of correlations at investment horizons before the correlations radically change pattern. This is visible in Figures 2–4 in the form of structural breaks (more details on structural breaks in assets are presented in Section 4.3).[12] For example, the gold-oil pair shows a significant increase in the overall correlation inferred from the DCC GARCH estimates during the periods 1994–1996 and 1998–2000. While DCC GARCH in fact shows the averaged correlation over the various investment horizons, wavelet correlations further reveal that this increase might be induced specifically by the long term correlations because correlations on short horizons are zero. We further observe that before the structural break, correlations between gold and stocks were very heterogeneous across various investment horizons implying a potential for risk diversification related to different investment horizons. However, after the structural break we can notice that correlations became very homogeneous in terms of their patterns (gold-oil, gold-stocks), which implies that gold, oil, and stocks could not be used effectively in one portfolio at the same time for risk diversification. This finding goes against the results of Baur and Lucey (2010), who find gold to be a good hedge against stocks and moreover a safe haven in extreme stock market conditions. In addition, Ciner et al. (2013) show that gold can be regarded as a safe haven against U.S. and UK exchange rates. In general, however, our result resonates well with the argument of Bartram and Bodnar (2009) that diversification provided little help to investors during the financial crisis.

The pattern described above can also be attributed to changes in investors' beliefs, which become homogeneous across investment horizons after the structural break.[13] This can be partially caused by broader uncertainty on financial markets; the influence of time-varying uncertainty on price formation and the diversification benefits is shown by Connolly et al. (2007) during 1992–2002 across a number of major stock markets. Another good reason for more homogeneous correlations across investment horizons could be the movement of investors away from passive investment strategies to more aggressive ones. Finally, after the introduction of full electronic trading on exchange platforms in 2005, the volume of automatic trading increased rapidly, a feature that may have increased homogeneity in correlations as well.

Thanks to the ability of wavelet analysis to provide results at different investment horizons, we are able to generalize inferences related to risk diversification. When correlations differ in their magnitudes at different investment horizons, risk diversification is possible. Low, negligible, or even no differences in correlation magnitudes at different investment horizons on the other hand preclude effective risk diversification in time. With the assets under research, risk diversification seems to be successful until 1991 for all three pairs.

---

[12] We formally tested for structural breaks in correlations by employing the supF test (Hansen, 1992; Andrews, 1993; Andrews and Ploberger, 1994) with p-values computed based on Hansen (1997).
[13] The importance of investors' beliefs and how they act on them has been recently shown by Ben-David and Hirshleifer (2012).

Later on differences emerge: (i) the diversification potential continues for the gold-stocks pair until the financial crisis eruption in 2007; (ii) diversification is limited from 1991 until the crisis for the gold-oil and oil-stocks pairs as correlations are largely homogenous; finally, (iii) no pair allows for diversification during the post-crisis period as correlations become homogenous.

*4.3 Structural changes and long-term equilibrium links: Cointegration*

Finally, we test for the existence of structural changes and when accounting for them, we test for long-term equilibrium relationships among assets. There exist reasons why gold and oil might share a cointegration relationship. Both assets are sensitive to strong economic and political events (Aggarwal and Lucey, 2007). The simple fact that both assets in our data set are quoted in U.S. dollars creates a potential link as well because volatile movements in the U.S. dollar would affect both assets in the same direction. Plus, Tully and Lucey (2007) show that the U.S. dollar is a key variable that exhibits a strong link towards gold and Zhang et al. (2008) show the same for oil. Further, gold is quite resistant with respect to inflation while increases in oil prices usually affect the aggregate price level and lead to an increase in inflation, which makes investment in gold more attractive. Finally, when major oil deliveries are paid during periods of higher oil prices, producers might use excess proceeds to buy gold whose price would increase due to higher demand (Zhang and Wei, 2010).

We have tested three pairs of assets for structural breaks and cointegration in the following way. First, we employed the sup*F* test (Hansen, 1992; Andrews, 1993; Andrews and Ploberger, 1994), with *p*-values computed based on Hansen (1997), and applied it to derived correlations between pairs of assets to endogenously search for the presence of structural breaks.[14] Hence, the endogenously detected break in the correlation series for a specific pair of assets is taken as a break to divide the sample of data to test for cointegration between the same pair of assets. Second, based on the results of the break test we divided the data into individual pre-break and post-break sub-samples that differ in their spans depending on the date of the identified break. For the asset pairs, the break dates were identified as follows: September 8, 2006 for gold-oil; May 5, 2009 for gold-stocks; and September 26, 2008 for oil-stocks. Finally we tested for cointegration between pairs of assets by employing the standard Johansen's procedure with appropriately chosen numbers of lags.[15] The cointegration results are presented in Table 5 for three periods: the pre-break period, the post-break period, and the full period (1987–2012).

For all three periods and all pairs of assets the result is unique: no cointegration was found. We take this result as support of our other findings. The absence of cointegration is in line with the low correlation found earlier during the period before the structural break. After all, if there is no long-term relationship between assets, why should there be a sizable correlation between them? Our results contradict earlier findings (Narayan et al., 2010; Zhang and Wei, 2010) but we credit this discrepancy to our use of high-frequency data, analysis at various investment horizons, and accounting for structural breaks. Further, the lack of cointegration is in line with a lack of economic linkages among these assets in terms of production, substitution, and complementary relationships as voiced by Casassus et al. (2013), and that underlines the long-term equilibrium relationship among assets or commodities. Finally, correlation and cointegration based on a long-term equilibrium might be sensitive to the selection of data frequency. However, it has been shown that markets that share a long-term equilibrium also exhibit a high degree of correlation on the intra-day frequency (Egert and Kočenda, 2011) and cointegration is often missing among less developed emerging markets (Egert and Kočenda, 2007).

---

[14] To conserve space, we do not report the test statistics for the detection of structural breaks. The results are available upon request.

[15] This result is robust with respect to lag selection as it does not change for any tested lag from 0 to 5. The results are available upon request.

## 5   Conclusions

In this paper we analyze dynamic correlations between pairs of key traded assets by employing a time-frequency approach with a wavelet methodology; the realized volatility and DCC GARCH approaches serve for comparison. In terms of the dynamic method the wavelet-based correlation analysis enables analyzing co-movements among assets not only from a time series perspective but also from the investment horizon perspective. Hence, we are able to provide unique evidence on how correlations between major assets vary over time and different investment horizons. We analyze the dynamic correlations of the prices of gold, oil, and the broad U.S. stock market index S&P 500 over 26 years from January 2, 1987 until December 31, 2012. The analysis is performed on both intra-day and daily data.

Our findings suggest a superior performance of wavelet analysis over standard benchmark approaches: wavelet analysis offers rich evidence of the heterogenous patterns in linkages among assets over time and across number of investment horizns. We show that heterogeneity in correlations across a number of investment horizons and between pairs of assets is a dominant feature during times of economic downturn and financial turbulence for all three pairs of asset under research. Moreover, heterogeneity prevails for most of the pre-crisis period in correlations between gold and stocks. The period when correlations between assets across investment horizons are homogenous run from the early 1990s until the financial crisis in 2008 (gold-oil, oil-stocks).

The post-crisis development is marked by a dramatically increased extent of correlations among all three assets and homogenous correlations as well. The timing of these changes differs for the three pairs. However, increases in correlations and their homogeneity occur after the structural breaks that have been identified in specific correlation series. After the breaks, the correlations for all pairs increased on average, but their magnitudes exhibited large swings up and down. Despite this strongly varying behavior, the correlations between pairs of assets became homogeneous and did not differ at distinct investment horizons. A strong implication emerges. During the period under research, and from different investment horizons perspectives, all three assets could be used in a well-diversified portfolio less often than common perception would have it.

**References**

Aggarwal, R. and Lucey, B.M. (2007). Psychological barriers in gold prices? Review of Financial Economics, 16, 217–230.

Aguiar-Conraria, L., M. M. Martins, and M. J. Soares (2012). The yield curve and the macro-economy across time and frequencies. Journal of Economic Dynamics and Control 36, 1950–1970.

Andersen, T. and L. Benzoni (2007). Realized volatility. In T. Andersen, R. Davis, J. Kreiss, and T. Mikosch (Eds.), Handbook of Financial Time Series. Springer Verlag.

Andersen, T., T. Bollerslev, F. Diebold, and P. Labys (2003). Modeling and forecasting realized volatility. Econometrica (71), 579–625.

Andrews, D. W. K. (1993). Tests for parameter instability and structural change with unknown change point. Econometrica, 61:821–856.

Andrews, D. W. K. and W. Ploberger (1994). Optimal tests when a nuisance parameter is present only under the alternative. Econometrica, 62:1383–1414.

Baffes, J., (2007). Oil spills on other commodities. Resources Policy 32, 126–134.

Bandi, F. and J. Russell (2006). Volatility. In J. Birge and V. Linetsky (Eds.), Handbook of Financial Engineering. Elsevier.

Barndorff-Nielsen, O. and N. Shephard (2004). Econometric analysis of realized covariation: High frequency based covariance, regression, and correlation in financial economics. Econometrica 72(3), 885–925.

Bartram, S.M, Bodnar, G.M., 2009. No place to hide: The global crisis in equity markets in 2008/2009. Journal of International Money and Finance, 28, 1246-1292.

Batten, J., Ciner, C., and Lucey, B.M. (2010). The Macroeconomic Determinants of Volatility in Precious Metals Markets, Resources Policy, 35 (2), 65-71.

Baur, D.G. and Lucey, B.M. (2010). 'Is Gold a Hedge or a Safe Haven? An Analysis of Stocks, Bonds and Gold', Financial Review, 45 (2), 217-29.

Bauwens, L., Laurent, S. (2005). A New Class of Multivariate Skew Densities, with Application to GARCH Models. Journal of Business and Economic Statistics 23(3): 346-354.

Bekaert, G., Hodrick, R.J., Zhang, X. (2009) International Stock Return Comovements, Journal of Finance, 64(6), 2591-2626.

Bekaert, G., Baele, L., Inghelbrecht, K. (2010). The Determinants of Stock and Bond Return Comovements. Review of Financial Studies, 23(6), 2374-2428.

Ben-David, I., Hirshleifer, D. (2012). Are Investors Really Reluctant to Realize their Losses? Trading Responses to Past Returns and the Disposition Effect, Review of Financial Studies, 25(8), 2485-2532.

Bollerslev, T. (1990). Modeling the coherence in short-run nominal modeling the coherence in short-run nominal exchange rates: A multivariate generalized arch model. Review of Economics and Statistics 72, 498–505.

Büyükşahin, B., Robe, M.A., 2013. Speculators, Commodities and Cross-Market Linkages, Journal of International Money and Finance, doi: 10.1016/j.jimonfin.2013.08.004.

Casassus, J., Liu, P., Tang, K., (2013). Economic Linkages, Relative Scarcity, and Commodity Futures Returns. Review of Financial Studies, 26(5): 1324-1362.

Ciner, C, Lucey, B and Gurdgiv, C (2013). Hedges and Safe Havens: An Examination of Stocks, Bonds, Gold, Oil and Exchange Rates. International Review of Financial Analysis, 29, 202–211.

Conlon, T., Cotter, J., Gençay, R. (2012). Commodity futures hedging, risk aversion and the hedging horizon. Geary Institute, University College Dublin, WP no. 2012/18.

Connolly, R. A., Stivers, C., and Sun, L. (2007). Commonality in the time-variation of stock-stock and stock-bond return comovements. Journal of Financial Markets, 10(2), 192-218.

Daubechies, I. (1988). Orthonormal bases of compactly supported wavelets. Communications on Pure and Applied Mathematics 41, 909-996.

Egert, B., Kočenda, E. (2007). Interdependence between Eastern and Western European Stock Markets: Evidence from Intraday Data. Economic Systems, 31(2), 184-203.

Egert, B., Kočenda, E. (2011). Time-Varying Synchronization of the European Stock Markets. Empirical Economics, 40(2), 393-407.

El-Sharif, Idris, Dick Brown, Bruce Burton, Bill Nixon, Alex Russell, (2005). Evidence on the nature and extent of the relationship between oil prices and equity values in the UK. Energy Economics 27, 819–830.

Engle, R. (2002). Dynamic conditional correlation: A simple class of multivariate generalized autoregressive conditional heteroskedasticity models. Journal of Business & Economic Statistics 20(3), 339–350.

Engle, R.F., Sheppard, K. (2001) Theoretical and Empirical Properties of Dynamic Conditional Correlation Multivariate GARCH. NBER Working Paper 8554.

Faff, R., Brailsford, T., (1999). Oil price risk and the Australian stock market. Journal of Energy Finance and Development 4, 69–87.

Faÿ, G., Moulines, E. Roueff, F. and Taqqu, M. S. (2009). Estimators of long-memory: Fourier versus wavelets. Journal of Econometrics 151 (2), 159 – 177.

Fernandez V. 2006. The impact of major global events on volatility shifts: evidence from the Asian crisis and 9/11. Economic Systems 30: 79–97.

Fernandez V. 2008. The war on terror and its impact on the long-term volatility of financial markets. International Review of Financial Analysis 17: 1–26.

Forbes, K.J., Rigobon, R. (2020). No Contagion, Only Interdependence: Measuring Stock Market Comovements. Journal of Finance, 57(5), 2223–2261.

Fratzscher, M., Schneider, D., Van Robays, I. (2013). Oil Prices, Exchange Rates and Asset Prices. *CESifo Working Paper* No. 4264.

Gadanecz, B. and Jayaram, K., (2009). Measures of financial stability - a review. Bank for International Settlements, IFC Bulletin No 31, 365-382.

Gallegati, M., M. Gallegati, J. B. Ramsey, and W. Semmler (2011). The US wage Phillips curve across frequencies and over time. Oxford Bulletin of Economics and Statistics 73(4), 489–508.

Gençay, R., Gradojevic, N., Selçuk, F., and Whitcher, B., (2010). Asymmetry of information flow between volatilities across time scales. Quantitative Finance, 10(8), 895-915.

Gençay, R., Selçuk, F., and Whitcher, B., (2001). An Introduction to Wavelets and Other Filtering Methods in Finance and Economics. San Diego, CA: Academic Press.

Graham, M., Kiviaho, J., and Nikkinen, J. (2013). Short-term and long-term dependencies of the S&P 500 index and commodity prices. Quantitative Finance, 13(4), 583-592.

Green, T.C., Hwang, B.H. (2009) Price-Based Return Comovement, Journal of Financial Economics 93, 37-50.

Greenwood, R. (2008). Excess Comovement of Stock Returns: Evidence from Cross-sectional Variation in Nikkei 225 Weights. Review of Financial Studies, 21(2), 1153–1186.

Hansen, P. and A. Lunde (2006). Realized variance and market microstructure noise. Journal of Business and Economic Statistics 24 (2), 127–161.

Hansen, B. E. (1992). Tests for parameter instability in regressions with I(1) processes. Journal of Business & Economic Statistics, 10:321–335.

Hansen, B. E. (1997). Approximate asymptotic p values for structural-change tests. Journal of Business & Economic Statistics, 15:60–67.

Huang, R.D., Masulis, R.W., Stoll, H.R. (1996). Energy shocks and financial markets. Journal of Futures Markets 16, 1 –27.

Hunt, B., (2006) Oil price shocks and the U.S. stagflation of the 1970s: Some insights from GEM, The Energy Journal, 27, 61-80.

In F, Kim S. 2006. The hedge ratio and the empirical relationship between the stock and futures markets: a new approach using wavelet analysis. Journal of Business 79: 799–820.

Kim S, In F. 2005. The relationship between stock returns and inflation: new evidence from wavelet analysis. Journal of Empirical Finance 12: 435–444.

Kim S, In F. 2007. On the relationship between changes in stock prices and bond yields in the G7 countries: wavelet analysis. Journal of International Financial Markets, Institutions and Money 17: 167–179.

Karuppiah J, Los C. 2005. Wavelet multiresolution analysis of high-frequency Asian FX rates, summer 1997. International Review of Financial Analysis 14: 211–246.

Khordagui, H. and Al-Ajmi, D. (1993). Environmental impact of the Gulf War: An integrated preliminary assessment. Environmental Management, 17(4), 557-562.

Lombardi, M.J. and I. Van Robays (2011): Do Financial Investors Destabilize the Oil Price? European Central Bank, Working Paper No.1346.

Lucey, B. Larkin, C. and O'Connor, F (2013). London or New York : Where and When does the gold price originate? Applied Economic Letters, 20(8), 813-817.

Marshall, John F. (1994). The Role of the Investment Horizon in Optimal Portfolio Sequencing (An Intuitive Demonstration in Discrete Time). Financial Review, 29(4), 557–576.

McAleer M, M. M. (2008). Realized volatility: A review. Econometric Reviews (27), 10–45.

Narayan, P.K., Narayan, S., and Zheng, X. (2010). Gold and oil futures markets: Are markets efficient? Applied Energy 87(10), 3299–3303.

Nekhili R, Altay-Salih A, Gençay R. 2002. Exploring exchange rate returns at different time horizons. Physica A 313: 671–682.

Nikkinen, J., Pynnönen, S., Ranta, M., Vähämaa, S. (2011). Cross-dynamics of exchange rate expectations: a wavelet analysis. International Journal of Finance & Economics, 16(3), 205–217.

Percival, D. B. (1995). On estimation of the wavelet variance. Biometrika 82, 619–631.

Percival, D. B. and A. T. Walden (2000). Wavelet Methods for Time series Analysis. Cambridge University Press.

Ramsey, J. B. (2002). Wavelets in economics and finance: Past and future. Studies in Nonlinear Dynamics & Econometrics 6(3).

Rua A, Nunes L. 2009. International comovement of stock market returns: a wavelet analysis. Journal of Empirical Finance 16: 632–639.

Sadorsky, P. (1999). Oil price shocks and stock market activity. Energy Economics 21, 449–469.

Sadorsky, P., (2001). Risk factors in stock returns of Canadian oil and gas companies. Energy Economics 23, 17– 28.

Samuelson, P.A. (1989). The judgement of economic science on rational portfolio management: Indexing, timing, and long-horizon effects. Journal of Portfolio Management, 16, 4-12.

Serroukh, A., A. T. Walden, and D. B. Percival (2000). Statistical properties and uses of the wavelet variance estimator for the scale analysis of time series. Journal of the American Statistical Association 95, 184–196.

Tully, Edel, and Brian M. Lucey, (2007). A power GARCH examination of the gold market. Research in International Business and Finance 21, 316–325.

Vacha, L. and J. Barunik (2012). Co-movement of energy commodities revisited: Evidence from wavelet coherence analysis. Energy Economics 34 (1), 241–247.

Whitcher, B., P. Guttorp, and D. B. Percival (1999). Mathematical background for wavelets estimators for cross covariance and cross correlation. Tech. Rep. 38, Natl. Res. Cent. for stat. and the Environ.

Whitcher, B., P. Guttorp, and D. B. Percival (2000). Wavelet analysis of covariance with application to atmosferic time series. Journal of Geophysical Research 105, 941–962.

Zhang, Y. and Wei, Y. (2010). The crude oil market and the gold market: Evidence for cointegration, causality and price discovery. Resources Policy 35, 168–177.

Zhang, Y., Fan, Y., Tsai, H., Wei, Y. (2008): Spillover Effect of US Dollar Exchange Rate on Oil Prices. Journal of Policy Modeling, 30, 973-991.

**Appendix A. Wavelet variance**

For a real-valued covariance stationary stochastic process $x(t)$, $t = 1,2, \ldots, N$, with mean zero, the sequence of the MODWT wavelet coefficients $W_x(j,s)$, for all $j, s > 0$ unaffected by the boundary conditions, obtained by the wavelet decomposition at scale $j$ is also a stationary process with mean zero. The wavelet variance at scale $j$ is the variance of wavelet coefficients at scale $j$, i.e.,

$$v_x(j)^2 = var(W_x(j,s)). \tag{A1}$$

For process $x(t)$, the estimator of the wavelet variance at level $j$ is defined as

$$\hat{v}_x(j)^2 = \frac{1}{M_j} \sum_{s=L_j-1}^{N-1} W_x(j,s)^2, \tag{A2}$$

where $M_j = N - L_j + 1 > 0$ is the number of $j$-th level MODWT coefficients that are unaffected by boundary conditions and $L_j$ denotes length of a wavelet filter at scale $j$ (Serroukh et al., 2000). While the variance of a covariance stationary process $x(t)$ is equal to the integral of the spectral density function $S_x(.)$, the wavelet variance at a particular level $j$ is the variance of the wavelet coefficients $W_x(j,s)$ with spectral density function $S_x(j)(.)$:

$$v_x(j)^2 = \int_{-1/2}^{1/2} S_x(j)(f) df = \int_{-1/2}^{1/2} \mathcal{H}_j(f) S_x(j)(f) df, \tag{A3}$$

where $\mathcal{H}_j(f)$ is the squared gain function of the wavelet filter $h_j$ (Percival and Walden, 2000). Since the variance of a process $x(t)$ is the sum of the contributions of the variances at all scales we can write:

$$var(x) = \sum_{j=1}^{\infty} v_x(j)^2. \tag{A4}$$

However, for a finite number of scales we have:

$$var(x) = \int_{-1/2}^{1/2} S_x(f) df = \sum_{j=1}^{J} v_x(j)^2 + var(V_x(J,s)). \tag{A5}$$

**Appendix B. Wavelet covariance**

Let $x(t)$ and $y(t)$ be covariance stationary processes with the square integrable spectral density functions $S_x(.)$, $S_y(.)$, and cross spectra $S_{xy}(.)$. Since we use an LA8 wavelet with length $L = 8$, we can use a generally non-stationary process, that is, it is stationary after the $d$-th difference, where $d \leq L/2$. The wavelet covariance of $x(t)$ and $y(t)$ at level $j$ is then defined as:

$$\gamma_{xy}(j) = Cov(W_x(j,s), W_y(j,s)). \tag{B1}$$

For a particular level of decomposition $J \leq log_2(N)$, the covariance of $x(t)$ and $y(t)$ is the sum of the covariances of the MODWT wavelet coefficients $\gamma_{xy}(j)$ at all scales $j = 1,2, \ldots, J$ and the covariance of the scaling coefficients $V_x(J,s)$ at scale $J$:

$$Cov(x_t, y_t) = Cov(V_x(J,s), V_y(J,s)) + \sum_{j=1}^{J} \gamma_{xy}(j). \tag{B2}$$

For the processes $x(t)$ and $y(t)$ defined above, the estimator of the wavelet covariance at level $j$ is defined as

$$\hat{\gamma}_{xy}(j) = \frac{1}{M_j} \sum_{s=L_j-1}^{N-1} W_x(j,s) W_y(j,s), \tag{B3}$$

where $M_j = N - L_j + 1 > 0$ is the number of $j$-th level MODWT coefficients for both processes that are unaffected by boundary conditions and $L_j$ denotes the length of the wavelet filter at scale $j$. Whitcher et al. (1999) prove that for Gaussian processes $x(t)$ and $y(t)$, the MODWT estimator of wavelet covariance is unbiased and asymptotically normally distributed.

# Tables

|  | High-frequency data | | | Daily data | | |
|---|---|---|---|---|---|---|
|  | **Gold** | **Oil** | **Stocks** | **Gold** | **Oil** | **Stocks** |
| Mean | 1.000E-06 | 3.185E-06 | -2.457E-06 | 2.216E-04 | 2.423E-04 | 2.699E-04 |
| St. dev. | 0.001 | 0.002 | 0.001 | 0.010 | 0.023 | 0.012 |
| Skewness | -0.714 | 1.065 | 0.326 | -0.147 | -1.063 | -0.392 |
| Kurtosis | 47.627 | 104.561 | 32.515 | 10.689 | 19.050 | 11.474 |
| Minimum | -0.042 | -0.045 | -0.024 | -0.077 | -0.384 | -0.098 |
| Maximum | 0.023 | 0.163 | 0.037 | 0.103 | 0.136 | 0.107 |

Table 1: Descriptive Statistics for high-frequency and daily gold, oil and, stock (S&P 500) returns over the sample period extending from January 2, 1987 until December 31, 2012.

|  | Gold-Oil | | | | | | | | | | |
|---|---|---|---|---|---|---|---|---|---|---|---|
|  | High-frequency data | | | | | Daily data | | | | | |
|  | 10 min | 20 min | 40 min | 80 min | 160 min-year | 2 days | 4 days | 8 days | 16 days | 32 days | 64 d. - year |
| **1987** | 0,02 | 0,03 | 0,08 | 0,15 | 0,80 | 0,12 | 0,00 | 0,14 | 0,07 | -0,13 | 0,77 |
| **1988** | 0,11 | 0,19 | 0,19 | 0,23 | 0,42 | 0,23 | 0,25 | 0,32 | 0,38 | 0,55 | 0,93 |
| 1989 | 0,02 | 0,03 | 0,06 | 0,06 | 0,53 | 0,02 | 0,11 | 0,05 | 0,09 | 0,74 | -0,14 |
| **1990** | 0,16 | 0,27 | 0,30 | 0,29 | 0,43 | 0,51 | 0,47 | 0,40 | 0,33 | 0,76 | 0,69 |
| **1991** | 0,21 | 0,32 | 0,31 | 0,32 | 0,63 | 0,01 | 0,00 | 0,47 | 0,48 | -0,31 | -0,47 |
| 1992 | 0,02 | 0,08 | 0,03 | 0,01 | -0,56 | 0,06 | 0,01 | -0,18 | 0,25 | -0,05 | -0,41 |
| 1993 | 0,00 | 0,01 | 0,04 | 0,02 | 0,60 | 0,12 | 0,04 | 0,20 | 0,30 | -0,26 | -0,77 |
| 1994 | 0,02 | 0,02 | 0,03 | 0,03 | -0,13 | 0,16 | 0,37 | 0,22 | -0,09 | -0,28 | 0,33 |
| 1995 | 0,01 | 0,00 | 0,03 | 0,07 | 0,05 | 0,23 | 0,17 | 0,07 | -0,02 | 0,16 | 0,39 |
| 1996 | 0,01 | 0,02 | 0,00 | 0,04 | -0,62 | -0,09 | -0,03 | 0,13 | -0,34 | -0,31 | -0,68 |
| 1997 | 0,00 | -0,01 | 0,00 | 0,06 | 0,33 | 0,00 | -0,22 | 0,04 | -0,13 | 0,09 | 0,57 |
| 1998 | 0,00 | -0,02 | -0,01 | -0,01 | 0,65 | 0,14 | 0,28 | 0,40 | 0,21 | 0,65 | 0,18 |
| 1999 | 0,01 | 0,01 | -0,01 | 0,02 | -0,58 | -0,02 | 0,12 | 0,31 | -0,17 | 0,17 | -0,80 |
| 2000 | 0,00 | 0,00 | 0,01 | 0,07 | -0,68 | 0,16 | 0,03 | 0,32 | -0,12 | 0,01 | 0,44 |
| 2001 | 0,00 | 0,01 | 0,01 | 0,02 | -0,83 | 0,23 | 0,04 | 0,11 | -0,25 | -0,10 | 0,10 |
| 2002 | -0,01 | -0,01 | 0,04 | 0,07 | 0,62 | 0,10 | 0,03 | -0,17 | 0,08 | -0,64 | 0,86 |
| 2003 | 0,01 | 0,02 | 0,04 | 0,06 | 0,68 | 0,24 | 0,05 | -0,08 | 0,12 | 0,47 | 0,54 |
| 2004 | 0,04 | 0,08 | 0,10 | 0,11 | 0,40 | 0,17 | 0,38 | 0,23 | 0,13 | -0,58 | -0,74 |
| 2005 | 0,01 | 0,07 | 0,09 | 0,11 | -0,42 | 0,08 | 0,07 | 0,22 | 0,42 | 0,27 | 0,40 |
| **2006** | 0,11 | 0,17 | 0,30 | 0,35 | 0,74 | 0,37 | 0,53 | 0,47 | 0,58 | 0,57 | 0,92 |
| **2007** | 0,26 | 0,30 | 0,33 | 0,35 | 0,29 | 0,49 | 0,38 | 0,07 | 0,41 | 0,42 | 0,48 |
| **2008** | 0,32 | 0,35 | 0,39 | 0,39 | 0,74 | 0,44 | 0,45 | 0,55 | 0,67 | 0,41 | 0,27 |
| 2009 | 0,19 | 0,21 | 0,22 | 0,22 | -0,21 | 0,19 | 0,20 | 0,53 | -0,03 | -0,12 | 0,45 |
| 2010 | 0,33 | 0,34 | 0,36 | 0,37 | -0,30 | 0,29 | 0,35 | 0,48 | 0,57 | 0,07 | -0,37 |
| 2011 | 0,26 | 0,27 | 0,31 | 0,29 | 0,22 | 0,20 | 0,18 | 0,20 | 0,37 | 0,62 | 0,72 |
| 2012 | 0,40 | 0,42 | 0,42 | 0,41 | -0,36 | 0,37 | 0,40 | 0,63 | 0,43 | -0,19 | 0,71 |

Table 2: Time-frequency correlation estimates for the gold – oil pair. The high-frequency set contains wavelet correlation estimates based on high-frequency data. The daily set contains wavelet correlation estimates based on daily data. Grey background represents years where the hypothesis of homogeneous correlations is rejected.

|  | Gold-Stocks | | | | | | | | | | |
|---|---|---|---|---|---|---|---|---|---|---|---|
|  | High-frequency data | | | | | Daily data | | | | | |
|  | 10 min | 20 min | 40 min | 80 min | 160 min-year | 2 days | 4 days | 8 days | 16 days | 32 days | 64 d. - year |
| **1987** | 0.05 | -0.04 | -0.11 | -0.22 | -0.54 | -0.22 | -0.22 | -0.24 | -0.39 | -0.58 | 0.64 |
| **1988** | -0.02 | -0.05 | -0.14 | -0.22 | -0.21 | -0.25 | 0.08 | -0.06 | -0.17 | -0.12 | -0.54 |
| **1989** | -0.03 | -0.11 | -0.19 | -0.15 | -0.59 | 0.03 | -0.26 | -0.23 | 0.06 | -0.67 | -0.92 |
| **1990** | -0.04 | -0.15 | -0.20 | -0.25 | -0.77 | -0.32 | -0.33 | -0.28 | -0.10 | -0.36 | -0.84 |
| 1991 | -0.01 | -0.07 | -0.10 | -0.09 | -0.55 | -0.16 | -0.14 | 0.11 | 0.18 | 0.14 | -0.59 |
| 1992 | -0.03 | -0.04 | -0.03 | -0.10 | 0.52 | 0.01 | -0.01 | -0.21 | -0.28 | 0.37 | 0.21 |
| **1993** | -0.02 | -0.06 | -0.10 | -0.13 | 0.49 | -0.24 | -0.15 | -0.23 | 0.03 | -0.12 | 0.43 |
| **1994** | -0.02 | -0.10 | -0.17 | -0.21 | 0.41 | -0.26 | -0.18 | 0.05 | 0.00 | 0.38 | 0.31 |
| 1995 | -0.01 | -0.04 | 0.01 | -0.04 | -0.37 | -0.17 | -0.06 | -0.02 | 0.01 | -0.39 | 0.71 |
| **1996** | -0.04 | -0.12 | -0.08 | -0.13 | -0.26 | -0.20 | -0.27 | -0.24 | 0.47 | 0.57 | -0.77 |
| 1997 | -0.03 | -0.06 | -0.07 | -0.11 | -0.49 | -0.15 | -0.17 | -0.26 | -0.05 | 0.03 | -0.93 |
| **1998** | -0.05 | -0.07 | -0.13 | -0.11 | 0.80 | -0.03 | 0.17 | 0.28 | -0.05 | 0.49 | 0.43 |
| 1999 | -0.01 | -0.01 | -0.04 | 0.01 | 0.54 | -0.03 | 0.10 | 0.05 | 0.15 | 0.45 | -0.75 |
| **2000** | -0.03 | -0.07 | -0.10 | -0.20 | 0.54 | -0.03 | 0.00 | 0.24 | 0.32 | -0.25 | -0.80 |
| 2001 | -0.01 | -0.01 | 0.00 | 0.01 | -0.49 | -0.24 | -0.11 | 0.07 | 0.04 | 0.16 | 0.43 |
| **2002** | -0.27 | -0.34 | -0.38 | -0.37 | -0.58 | -0.21 | -0.24 | -0.37 | -0.28 | -0.34 | -0.66 |
| **2003** | -0.26 | -0.35 | -0.38 | -0.42 | 0.46 | -0.42 | -0.12 | -0.07 | -0.51 | -0.49 | 0.18 |
| 2004 | -0.07 | -0.09 | -0.08 | -0.09 | 0.65 | 0.03 | 0.14 | 0.38 | 0.14 | 0.17 | 0.29 |
| 2005 | -0.02 | -0.02 | 0.02 | 0.00 | 0.08 | -0.08 | 0.11 | 0.09 | -0.02 | 0.40 | 0.13 |
| **2006** | 0.05 | 0.11 | 0.17 | 0.20 | 0.30 | 0.10 | -0.01 | 0.20 | 0.34 | 0.65 | 0.16 |
| **2007** | 0.20 | 0.26 | 0.29 | 0.27 | -0.18 | 0.39 | 0.28 | 0.42 | 0.42 | 0.85 | 0.39 |
| 2008 | 0.11 | 0.14 | 0.10 | 0.09 | 0.87 | -0.03 | -0.16 | -0.09 | -0.16 | -0.68 | -0.89 |
| 2009 | 0.14 | 0.13 | 0.15 | 0.17 | -0.17 | 0.01 | -0.05 | 0.28 | 0.31 | -0.05 | -0.38 |
| 2010 | 0.25 | 0.25 | 0.28 | 0.29 | 0.06 | 0.14 | 0.29 | 0.46 | 0.38 | -0.08 | -0.20 |
| 2011 | 0.13 | 0.14 | 0.18 | 0.13 | -0.40 | -0.17 | -0.18 | -0.08 | -0.04 | 0.20 | 0.49 |
| 2012 | 0.40 | 0.39 | 0.37 | 0.38 | 0.67 | 0.42 | 0.26 | 0.62 | 0.58 | 0.06 | 0.05 |

Table 3: Time-frequency correlation estimates for the gold – stocks pair. The high-frequency set contains wavelet correlation estimates based on high-frequency data. The daily set contains wavelet correlation estimates based on daily data. Grey background represents years where the hypothesis of homogeneous correlations is rejected.

|  | Oil-Stocks | | | | | | | | | | |
| --- | --- | --- | --- | --- | --- | --- | --- | --- | --- | --- | --- |
|  | High-frequency data | | | | | Daily data | | | | | |
|  | 10 min | 20 min | 40 min | 80 min | 160 min-year | 2 days | 4 days | 8 days | 16 days | 32 days | 64 d. - year |
| **1987** | 0.03 | 0.01 | 0.05 | 0.04 | -0.64 | -0.11 | 0.21 | -0.07 | -0.09 | -0.03 | 0.66 |
| **1988** | 0.03 | -0.03 | -0.05 | -0.11 | 0.30 | -0.06 | 0.13 | -0.09 | -0.15 | -0.30 | -0.74 |
| **1989** | 0.01 | 0.00 | 0.03 | -0.02 | 0.13 | -0.08 | 0.12 | -0.06 | -0.10 | -0.74 | 0.06 |
| **1990** | -0.02 | -0.12 | -0.18 | -0.20 | -0.54 | -0.38 | -0.46 | -0.62 | -0.25 | 0.04 | -0.84 |
| **1991** | -0.04 | -0.10 | -0.17 | -0.19 | -0.49 | -0.06 | -0.06 | 0.38 | 0.34 | -0.51 | 0.58 |
| **1992** | 0.03 | 0.01 | 0.04 | -0.02 | -0.51 | 0.08 | 0.04 | 0.20 | -0.20 | -0.52 | 0.50 |
| **1993** | 0.00 | 0.00 | -0.01 | -0.02 | 0.73 | -0.05 | -0.11 | 0.24 | 0.42 | 0.10 | -0.63 |
| **1994** | 0.00 | 0.01 | -0.03 | -0.04 | -0.77 | -0.23 | -0.05 | 0.01 | 0.13 | -0.47 | -0.58 |
| **1995** | -0.02 | 0.01 | 0.02 | 0.00 | -0.47 | -0.05 | 0.04 | 0.06 | 0.37 | -0.14 | -0.20 |
| **1996** | 0.00 | 0.00 | 0.00 | -0.03 | -0.02 | -0.02 | 0.05 | -0.18 | -0.15 | -0.38 | 0.56 |
| **1997** | 0.00 | 0.00 | 0.07 | 0.08 | -0.61 | -0.15 | 0.00 | 0.02 | 0.08 | 0.19 | -0.50 |
| **1998** | 0.00 | -0.02 | 0.02 | 0.02 | 0.64 | 0.04 | 0.13 | 0.15 | 0.21 | 0.52 | -0.79 |
| **1999** | -0.01 | 0.02 | 0.03 | 0.00 | -0.94 | -0.03 | 0.01 | 0.09 | 0.17 | 0.73 | 0.99 |
| **2000** | 0.02 | -0.01 | -0.02 | -0.05 | -0.18 | -0.11 | -0.09 | 0.07 | -0.11 | -0.53 | -0.24 |
| **2001** | 0.01 | 0.04 | 0.03 | -0.05 | 0.51 | -0.12 | -0.04 | 0.08 | 0.03 | 0.84 | 0.81 |
| **2002** | -0.01 | -0.01 | -0.02 | -0.03 | -0.70 | 0.17 | 0.19 | 0.41 | 0.24 | 0.36 | -0.53 |
| **2003** | -0.01 | -0.03 | -0.04 | -0.05 | 0.57 | -0.24 | 0.08 | -0.49 | -0.36 | -0.58 | -0.54 |
| **2004** | -0.07 | -0.15 | -0.18 | -0.25 | 0.13 | -0.13 | 0.01 | 0.05 | -0.08 | -0.68 | -0.71 |
| **2005** | -0.16 | -0.19 | -0.17 | -0.18 | 0.00 | -0.09 | 0.14 | 0.04 | -0.46 | -0.20 | 0.32 |
| **2006** | 0.04 | 0.07 | 0.09 | 0.10 | -0.08 | 0.07 | 0.07 | 0.08 | 0.24 | 0.46 | -0.12 |
| **2007** | 0.09 | 0.13 | 0.13 | 0.12 | -0.63 | 0.17 | 0.07 | -0.04 | 0.04 | 0.18 | 0.74 |
| **2008** | 0.26 | 0.27 | 0.31 | 0.33 | 0.70 | 0.42 | 0.32 | 0.09 | 0.05 | 0.12 | -0.47 |
| **2009** | 0.42 | 0.46 | 0.48 | 0.50 | 0.92 | 0.53 | 0.28 | 0.61 | 0.10 | -0.05 | 0.51 |
| **2010** | 0.57 | 0.59 | 0.58 | 0.62 | 0.69 | 0.70 | 0.71 | 0.51 | 0.58 | 0.91 | 0.86 |
| **2011** | 0.50 | 0.53 | 0.53 | 0.56 | 0.32 | 0.53 | 0.57 | 0.62 | 0.53 | -0.03 | 0.74 |
| **2012** | 0.49 | 0.48 | 0.47 | 0.46 | 0.26 | 0.52 | 0.54 | 0.74 | 0.53 | 0.12 | 0.30 |

Table 4: Time-frequency correlation estimates for the oil – stocks pair. The high-frequency set contains wavelet correlation estimates based on high-frequency data. The daily set contains wavelet correlation estimates based on daily data. Grey background represents years where the hypothesis of homogeneous correlations is rejected.

|             | gold-oil |        | gold-stocks |        | oil-stocks |        |
|-------------|----------|--------|-------------|--------|------------|--------|
| pre-break   | 10.07    | (0.31) | 6.70        | (0.64) | 7.87       | (0.52) |
| post-break  | 9.75     | (0.34) | 7.66        | (0.55) | 6.94       | (0.62) |
| full period | 4.66     | (0.84) | 10.64       | (0.25) | 11.19      | (0.20) |

Table 5: Johansen's cointegration results. The table reports trace test statistics together with p-values in parentheses. The break dates dividing the period into pre-break and post-break are September 8, 2006, May 5, 2009, and September 26, 2008 for the gold-oil, gold-stocks and oil-stocks pairs, respectively. The full period covers January 2, 1987 to December 31, 2012.

# Figures

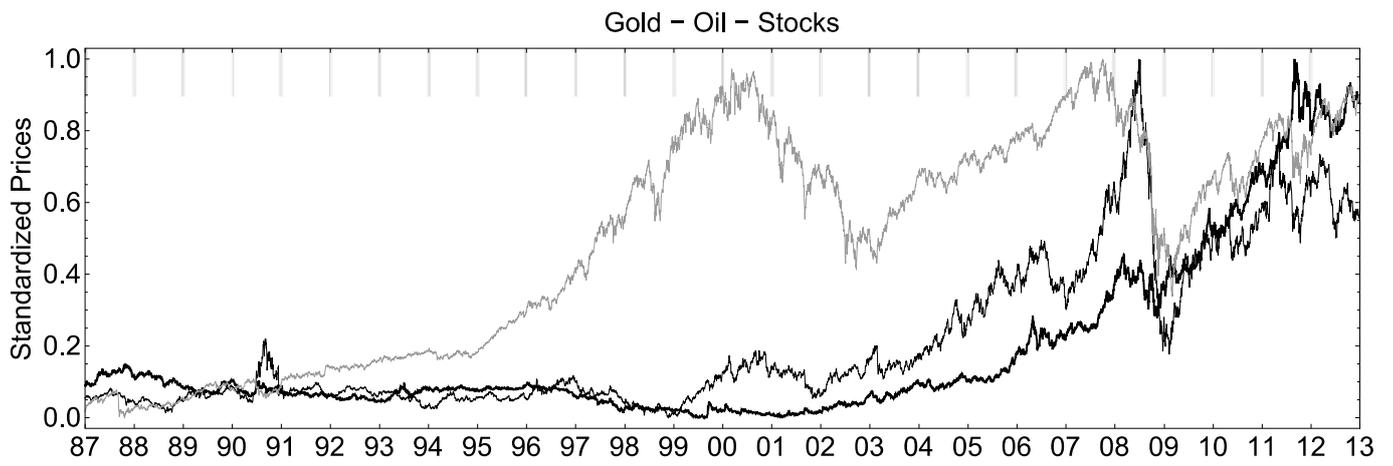

**Figure 1: Normalized prices of gold (thin black), oil (black), and stocks (gray).**

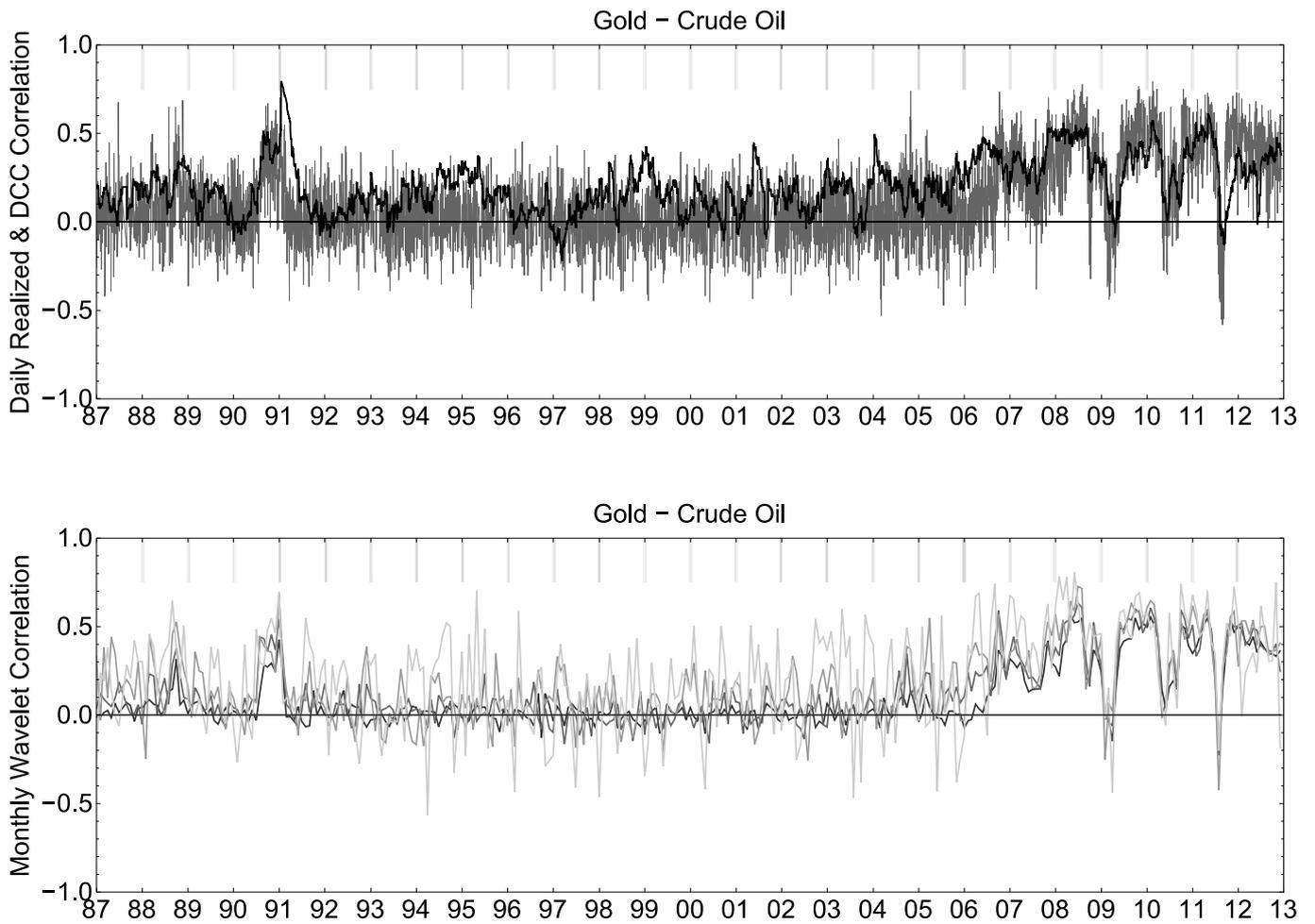

Figure 2: Dynamics in gold-oil correlations. The upper panel contains the realized correlation for each day of the sample (grey line) and daily correlations estimated from the DCC GARCH model (black line). The lower panel contains time-frequency correlations based on the wavelet correlation estimates from high-frequency data for each month separately. We report correlation dynamics at 10-minutes, 40-minute, 2.66 hour (approximate), and 1.6-day (approximate) horizons depicted by the thick black to thin black lines.

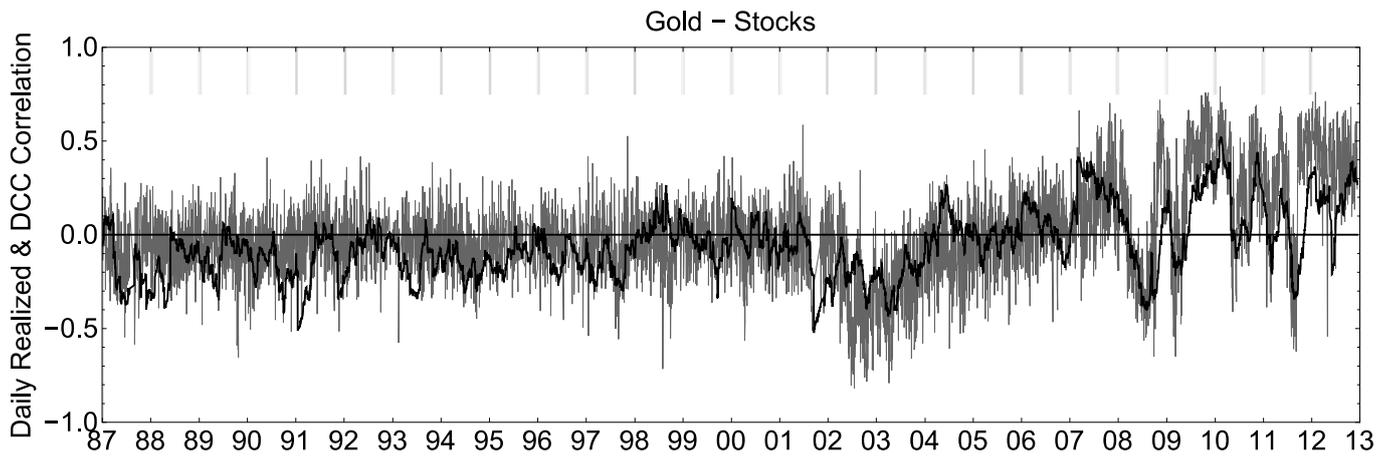

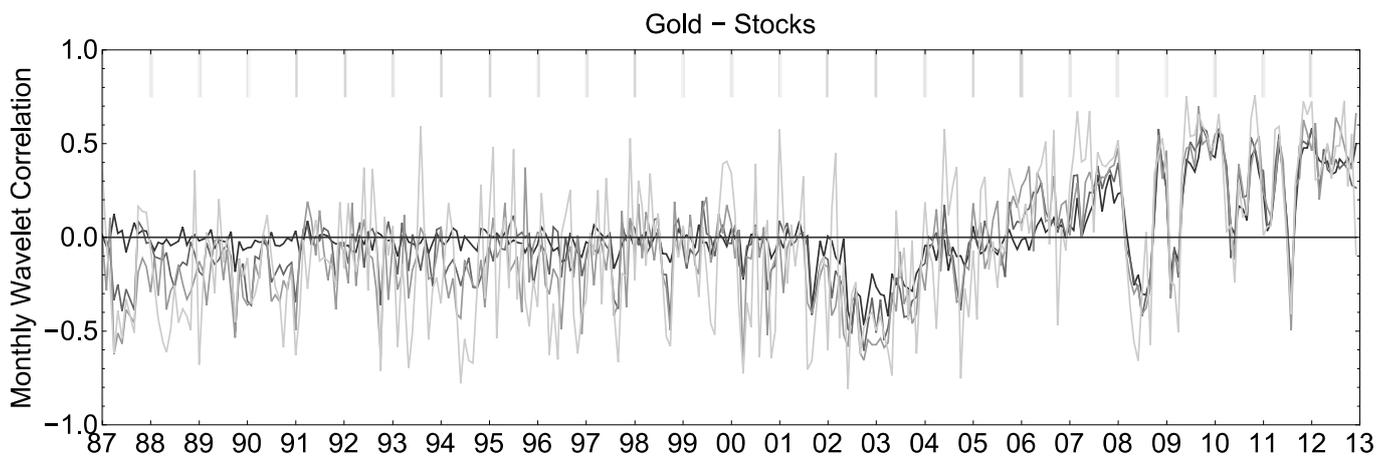

Figure 3: Dynamics in gold – stocks correlations. The upper panel contains the realized correlation for each day of the sample (grey line) and daily correlations estimated from the DCC GARCH model (black line). The lower panel contains time-frequency correlations based on the wavelet correlation estimates from high-frequency data for each month separately. We report correlation dynamics at 10-minute, 40-minute, 2.66 hour (approximate), and 1.6-day (approximate) horizons depicted by the thick black to thin black lines.

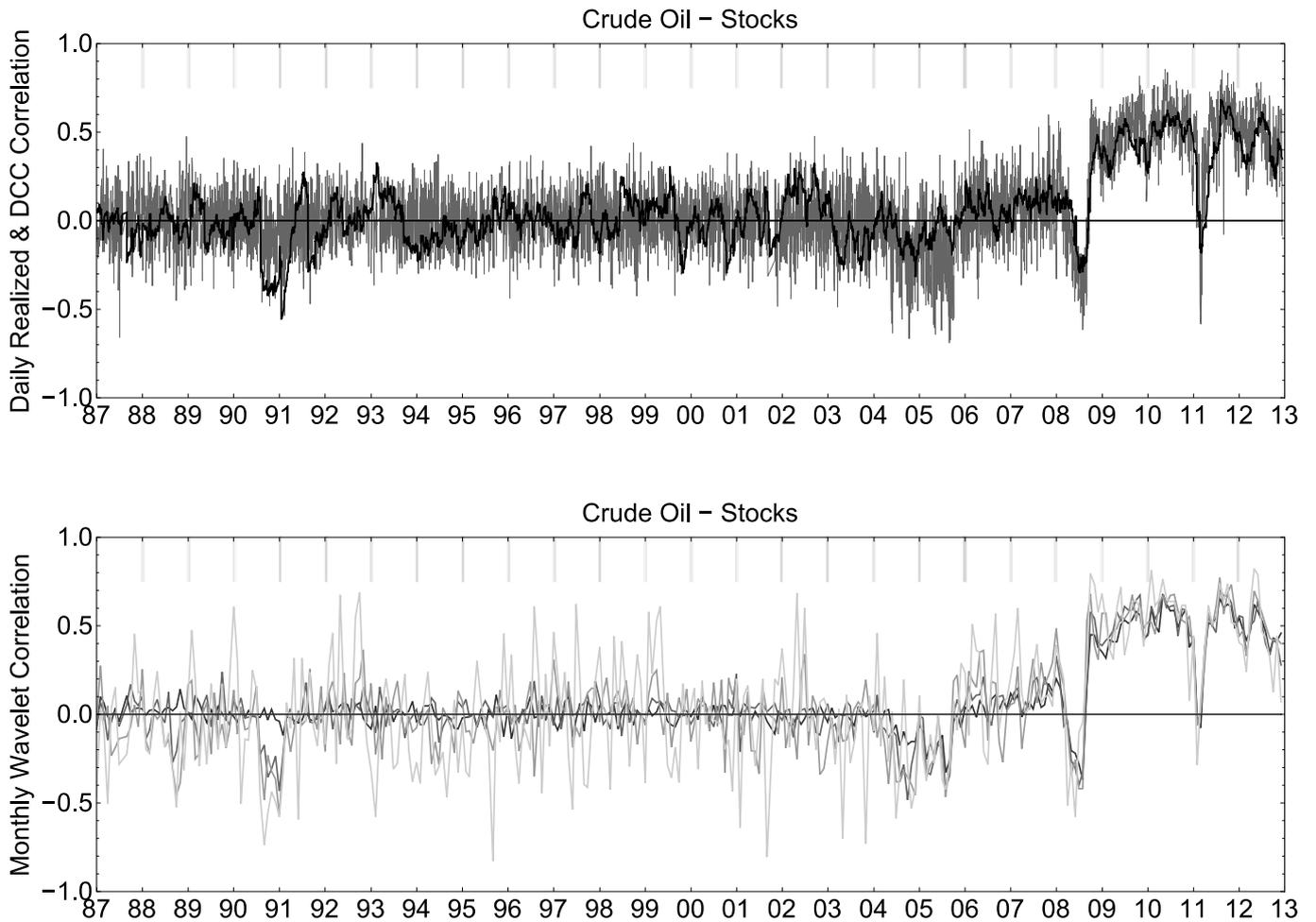

Figure 4: Dynamics in oil – stocks correlations. The upper panel contains the realized correlation for each day of the sample (grey line) and daily correlations estimated from the DCC GARCH model (black line). The lower panel contains time-frequency correlations based on the wavelet correlation estimates from high-frequency data for each month separately. We report correlation dynamics at 10-minute, 40-minute, 2.66 hour (approximate), and 1.6-day (approximate) horizons depicted by the thick black to thin black lines.